 \documentclass[preprint,review,12pt]{elsarticle}

\usepackage{amssymb}
\journal{Phys. Lett. A}

\begin{document}

\begin{frontmatter}



\title{Acoustic phonons mediated non-equilibrium spin current in the presence of Rashba and Dresselhaus spin-orbit couplings}

 \author{K. Hasanirokh}
 \author{ A. Phirouznia \corref{cor1}}
 \address{Department of Physics, Azarbaijan Shahid Madani University, 53714-161, Tabriz, Iran}

%\author{}

\begin{abstract}
Influence of electrons interaction with longitudinal acoustic
phonons on magnetoeletric and spin-related transport effects are
investigated. The considered system is a two dimensional electron
gas system with both Rashba and Dresselhaus spin-orbit couplings.
The works which have previously been performed in this field, have
revealed that the Rashba and Dresselhaus couplings cannot be
responsible for spin-current in the non-equilibrium regime. In the
current Letter, a semi-classical method was employed using the
Boltzmann approach and it was shown that the spin-current of the
system, in general, does not go all the way to zero when the
electron-phonon coupling is taken into account. It was also shown
that spin accumulation of the system could be influenced by
electron-phonon coupling.
\end{abstract}

\cortext[cor1]{Corresponding author. Tel.: +98-412-4327500, Fax:
+98-412-4327541} \ead{Phirouznia@azaruniv.ac.ir}


\end{frontmatter}

% \linenumbers

\section{\textbf{Introduction}}
\label{}

Spintronics has attracted more and more attention, from both
theoretical and experimental sides, during the last several years.
Effective control of spin polarized transport is very important,
especially for practical applications; for example, in multilayers.
Therefore, many studies have been conducted to explain this
phenomenon \cite{Hammar,Li}.
\\
According to the results of these studies, manipulation of spin, can
be realized by applying magnetic fields or Rashba interaction. The
Rashba interaction arises from inversion asymmetry in the system and
can be effectively controlled by applying a gate voltage
\cite{Rashba1,Rashba2,Bychkov}. Due to the tunable nature of Rashba
coupling, it is the most popular method for manipulation of electron
spin. This kind of spin-orbit interaction (SOI) plays a central role
in Datta and Das spin field-effect transistor (SFET) \cite{Datta}.
Generally, SOI has a significant role in magnetoresistance effects,
known as weak localization \cite{Bergmann}. In the field of
spin-transport, many interesting features have been demonstrated for
this type of spin-orbit coupling (SOC)
\cite{Lyanda,Iordanskii,Lyanda2}. For example, it has been verified
that spin-orbit scattering can induce localization/antilocalization
transition in a two-dimensional electron gas (2DEG) system
\cite{Miller,Koga}. Meanwhile, Rashba interaction has been also
suggested for spin interference devices and spin-filters
\cite{Aronov,Koga2,Kiselev}.
\\
Another spin-orbit coupling that provides a new parameter which
should be considered in designing spin-dependent devices, is the
Dresselhaus coupling. Dresselhaus coupling is induced by the bulk
inversion asymmetry \cite{Dresselhaus}. Effects of spin-orbit
couplings (SOCs) in semiconductors have attracted growing interest
due to their roles in semiconductor spintronics.
\\
Manipulation of spin makes new functionality in electronic devices.
Control of spin accumulation by spin-orbit interactions has a great
potential in the field of spintronics
\cite{Hammar4,Monzon,Zhian,Dimitrie}. In the presence of these two
different spin-orbit interactions, i.e. the Rashba and Dresselhaus
couplings, in a two dimensional electron gas system, one can
effectively control both magnitude and direction of non-equilibrium
spin accumulation\cite{Zhian}. Meanwhile, spin-current vanishes
exactly in the non-equilibrium regime induced by an in-plane driving
electric field \cite{Zhian}. This implies that there is no
spin-polarized current accompanied by the spin accumulation of the
system \cite{Zhian}.
\\
Rashba obtained non-vanishing spin-current in equilibrium state.
Therefore, this spin-current can not describe any real transport of
spins in non-equilibrium regime induced by an in-plane driving
electric field \cite{Rashba}. For a non-equilibrium system, a highly
anisotropic spin response to an in-plane electric field has been
discovered \cite{Maxim}. However, as mentioned before, based on the
semiclassical approach, Huang and Hu showed that the non-equilibrium
spin current vanishes exactly in two-dimensional electron gases, in
the presence of both Rashba and Dresselhaus couplings. Meanwhile
non-equilibrium spin accumulation can be obtained in this case
\cite{Zhian}. In addition, Inoue et al. obtained similar results
based on the Green's function approach for two-dimensional electron
gases \cite{Inue}.
\\
It should be noted that the work described by Huang and Hu
\cite{Zhian}, has been based on a semi-classical approach developed
by Schliemann and Loss \cite{Loss1}. They formulated the anisotropic
effects of energy dispersion relation and scattering matrixes in the
presence of spin-orbit couplings. Meanwhile, the exact solution to
the Boltzmann equation for two-dimensional anisotropic systems was
provided by V\'{y}born\'{y} et al. \cite{Karel}, in which it was
shown that, for a Rashba type two-band model, discrepancy between
 exact and approximate Schliemann and Loss approach, remains
only at the level of higher order corrections \cite{Karel}.
Therefore, the Schliemann and Loss approach was applied so that the
results of the current letter could be compared with the results of
 Huang and Hu \cite{Zhian} in the same theoretical framework. It
can be easily shown that, in the presence of both spin-orbit
couplings i.e. Rashba and Dresselhaus interactions, Schliemann and
Loss method is still a good approximation in comparison with the
exact V\'{y}born\'{y} approach, which could be due to the fact that
anisotropic term of the energy dispersion relation (which was
induced by Dresselhaus coupling) is negligible and the anisotropic
effects can be entered only through the scattering matrixes.
\\
The polaronic properties have been studied theoretically in the
presence of the Rashba and Dresselhaus spin-orbit interactions and
weak electron-phonon coupling in a pioneering work \cite{Vardanyan}.
It was demonstrated that in the presence of both spin-orbit
couplings, self-energy correction of the electron energy and polaron
effective mass show an angular anisotropy \cite{Vardanyan}. In this
letter, based on the mentioned semiclassical approach, influence of
the electron-phonon scattering has been considered on spin-transport
quantities in a two-dimensional electron gas when the spin-orbit
interactions are present. It was verified that the electron-phonon
scattering results in non-vanishing spin-current that could be
controlled by spin-orbit interactions. It has also been found that
the spin-current is influenced by the electron-phonon coupling. The
details of the numerical results have briefly been addressed in the
present letter.
\section{Model and approach}
The total Hamiltonian is given by
\begin{equation}
\hat{H}=\hat{H}_{0}+\hat{V}_{im} +\hat{H}_{el-ph},
\end{equation}
in which,~$\hat{H}_{0}$  is the kinetic energy and spin-orbit
interactions (including both Rashba and Dresselhaus spin-orbit
couplings), for a 2DEG namely
\begin{equation} \hat{H}_{0}=\frac{\hbar^2
k^2}{2m}+\alpha(\hat{\sigma}_x k_y -\hat{\sigma}_y
k_x)+\beta(\hat{\sigma}_x k_x-\hat{\sigma}_y k_y),
\end{equation}
where ${\bf k}$ is the wave vector of conduction electrons,
$\sigma_i (i=x,y)$ are Pauli matrices, $\alpha $  and $ \beta $
denote the strengths of the Rashba and Dresselhaus interaction,
respectively.
\\
For a given wave vector ${\bf k}$,
\begin{equation} \mid {\bf k}\lambda > = \frac{1}{\sqrt{2}}\left(
\begin{array}{cccc}
e^{i\frac{\phi_{\bf k}}{2}}\\
\lambda e^{-i\frac{\phi_{\bf k}}{2}}\\
\end{array}
\right),
\end{equation}
are the eigenfunctions of $\hat{H}_{0}$ where $\lambda=\pm1 $ and
$\phi_{\bf k}$ defined as
\begin{equation}
\tan\phi_{\bf k}=\frac{\alpha k_x+\beta k_y}{\alpha k_y+\beta k_x}.
\end{equation}
The corresponding eigenvalues of $\hat{H}_{0}$ are
\begin{equation}
\epsilon_{{\bf k}\lambda}=\frac{\hbar^2k^2}{2m}+\lambda
\sqrt{(\alpha^2+\beta^2)k^2+4\alpha\beta k_x k_y}.
\end{equation}
The expectation value of spin of electrons along the x and y
directions in a given state $\mid{\bf k}\lambda>$ can easily be
found as follows
\begin{equation}
S^{(0)}_{\lambda,x}({\bf k})=\frac{\hbar}{2} \lambda \cos(\phi_{\bf
k}),~~~ S^{(0)}_{\lambda,y}({\bf k})=\frac{-\hbar}{2} \lambda
\sin(\phi_{\bf k}).
\end{equation}
\\
The second term of Hamiltonian, $V_{im}(r)$ is potential of
impurities given as
\begin{equation}
V_{im}(r)=\sum_{i}(J\hat{\sigma}.\vec{m}(\textbf{r}))\delta({\bf
r}-{\bf r_{i}}),
\end{equation}
where the sum is performed  over all of the randomly distributed
impurities, J is the exchange interaction strength of magnetic
impurities with conduction electrons and $\hat{m}(\textbf{r})$ is
the unit vector along the local magnetization.
\begin{eqnarray}
<{\bf k}'\lambda'n\acute{_q}\mid V_{im}(r)\mid {\bf k}\lambda n_q>
=C_{{\bf k'},{\bf k}}\delta_{\acute{n}_q,n_q} \left(
\begin{array}{cccc}
+J_z && J_x-iJ_y\\
J_x+iJ_y && -J_z\\
\end{array}
\right).
\end{eqnarray}
In which
\begin{equation}
J_z=m_zJ,\;\;\;\;J_y=m_yJ,~~~~~ J_x=m_xJ,
\end{equation}
and
\begin{equation}
C_{{\bf k'},{\bf
k}}=(1/\sqrt{L_xL_y})\sum_{j}\exp(i(\vec{k'}-\vec{k}).\vec{r}_j).
\end{equation}
For long range magnetic interactions, because of the shape
anisotropy, we take $m_z=0$ and for randomly oriented magnetic
moments of impurities, one can assume
\begin{equation} < m_x > = <
m_y > =0,   ~~ < m^2_x > =< m^2_y >=\frac{1}{2}.
\end{equation}
Since it was assumed that the magnetic moments of the impurities
have been randomly oriented, therefore they can not be responsible
for spin-polarized effects and spin current.
\\
 The last term of the
Hamiltonian, $ \hat{H}_{el-ph}$ is electron-phonon interaction and
can be expressed as \cite{Hamaguchi},
\begin{equation}
\hat{H}_{el-ph}=D_{ac}\nabla.\vec{u}(r),
\end{equation}
here, $D_{ac}$ is defined as deformation potential for electron
scattering by acoustic phonons and
$\vec{u}(r)$  is a small displacement vector of an ion from its equilibrium position, $\vec{R}$.\\
For a two-dimensional system, the displacement is determined as
\begin{eqnarray}
\label{u}
\vec{u}(r)=\sum_{q}\sqrt{\frac{\hbar}{2MN\omega_q}}\hat{e_q}[a_q
e^{i\vec{q}.\vec{r}}+ a_q^\dag e^{-i\vec{q}.\vec{r}}],
\end{eqnarray}
where, M and N are mass and number of the ions, respectively.
$\hat{e_q}$ is a unit vector in displacement direction and
$\omega_q=V_sq$ in which $V_s$ is the sound velocity and $q$ is wave
vector of phonon.
\\
Using  from eq. (\ref{u}), the electron-phonon interaction can be
written as
\begin{eqnarray}
\hat{H}_{el-ph}=D_{ac}\sum_{q}\sqrt{\frac{\hbar}{2MNW_q}}(i\hat{e_q}.\vec{q})
[a_q e^{i\vec{q}.\vec{r}}- a_q^\dag e^{-i\vec{q}.\vec{r}}].
\end{eqnarray}
by defining
\begin{equation}
c(q)=D_{ac}\sqrt{\frac{\hbar}{2MNW_q}}(i\hat{e_q}.\vec{q}),
\end{equation}
we obtain the following result, directly
\begin{equation}
\hat{H}_{el-ph}=\sum_{q}[c(q)a_q e^{i\vec{q}.\vec{r}}+c^*(q)
a_q^\dag e^{-i\vec{q}.\vec{r}}].
\end{equation}
The eigenstate of the phonon Hamiltonian in harmonic approximation
is defined by $| n_q> $, where $n_q$ is the phonon occupation
number, so we can define a new basis as follows, $\mid {\bf
k}\lambda n_q>=\mid {\bf k}\lambda
>\otimes\mid n_q> $. Scattering matrix of electron-phonon
interaction is given as follows
\begin{eqnarray}
{<{{\bf k}'}{\lambda'}{{n'}_q}\mid {\hat{H}}_{el-ph}\mid {\bf
k}\lambda n_q>}=\left \{\begin{array}{ll}
\delta_{n'_q,n_q-1}\delta_{\lambda'\lambda}c(q)\sqrt{n_q}, & {\rm if}\ {\bf k}'={\bf k}+{\bf q} ,\\
\\
\delta_{n'_q,n_q+1}\delta_{\lambda'\lambda}c^*(q)\sqrt{n_q+1}, &
{\rm if}\ {\bf k}'={\bf k}-{\bf q} .\end{array} \right.
\end{eqnarray}
As mentioned, in the current letter we have employed the procedure
which have been used in \cite{Zhian}. The two last terms of the
Hamiltonian are responsible for both spin-dependent and
spin-independent relaxation mechanisms. If we rename these two terms
as
\begin{equation}
V=\hat{V}_{im}+\hat{H}_{el-ph}\nonumber
\end{equation}
The Lippman- Schwinger scattering state of a conduction electron
reads,
\begin{eqnarray}
\mid {\bf k}\lambda n_q>_{scat}&=&\mid {\bf k}\lambda n_q>
+\sum_{{\bf k'} q'\lambda' }\frac{V_{\bf k' \lambda' n'_q,\bf k
\lambda n_q}}{\epsilon_{{\bf k}\lambda}-\epsilon_{{\bf
k'}\lambda'}+i\eta}\mid {\bf k'}q'\lambda'>,
\end{eqnarray}
where $\eta$ is a small positive quantity. Then the spin expectation
value in a given scattering state is
\begin{eqnarray}
S_{\lambda,i}({\bf k})= S^{(0)}_{\lambda,i}({\bf k})
+\hbar\sum_{{\bf
k}'q'\lambda'}Re[\frac{V_{k'\lambda'\acute{n}_q,{\bf k}\lambda
n_q}<\sigma_i>_{{\bf k}'\lambda'\acute{n}_q,{\bf k}\lambda
n_q}}{\epsilon_{{\bf k}\lambda}-\epsilon_{{\bf k}'\lambda'}+i\eta}].
\end{eqnarray}
Here we have defined $S_{\lambda,i}({\bf k})= <{\bf k}\lambda n_q
\mid \hat{S}_i \mid {\bf k}\lambda n_q>_{scat}$. Therefore one can
obtain spin expectation value as seen in eq. (\ref{s}).
\begin{eqnarray}
\label{s} S_{\lambda,i}({\bf k})&=&S^{(0)}_{\lambda,i}({\bf k})
+\hbar\sum_{{\bf k}'q'\lambda'} Re[V_{{\bf
k}'\lambda'\acute{n}_q,{\bf k}\lambda n_q}<\hat{\sigma}_i>_{{\bf
k}'\lambda'\acute{n}_q,{\bf k}\lambda n_q} Pr\frac{1}{\epsilon_{{\bf
k}\lambda}-\epsilon_{{\bf
k}'\lambda'}} \nonumber\\
&&-V_{k'\lambda'\acute{n}_q,{\bf k}\lambda n_q} <\sigma_i>_{{\bf
k}'\lambda'\acute{n}_q,k\lambda
n_q}i\pi\delta(\epsilon_{k\lambda}-\epsilon_{k'\lambda'})].
\label{eq:wideeq}
\end{eqnarray}
Here $ <\sigma_i>_{{\bf k}'\lambda',{\bf k}\lambda}$ is the
expectation value of the Pauli Matrix for a Lippman- Schwinger
scattering state. Then, net spin density can be giving by
\begin{equation}
<S_i>=\sum_{{\bf k},q\lambda}S_{\lambda,i}({\bf k})f_\lambda({\bf
k},{\bf q}),
\end{equation}
in which $f_\lambda({\bf k},{\bf q})$ is non-equilibrium
distribution function of conduction electrons. Where in the absence
of external electric field, this can be reduced to the equilibrium
Fermi-Dirac distribution,
\begin{equation}
f_\lambda({\bf k},{\bf q})=f_0(\epsilon_{{\bf
k}\lambda})=\frac{1}{1+e^\frac{(\epsilon_{{\bf
k}\lambda}-\epsilon_F)}{k_BT}}.
\end{equation}
We have used the Debye model, so the summation over ${\bf q}$ is
easily calculated by replacing it with an integral. This integral
can be considered to be evaluated in an interval starting from ${\bf
q}=0$ to the Debye wave vector, ${\bf q}_D $. This wave vector is
directly related to the free-electron Fermi wave vector. In
two-dimensional metals, $q_D=\sqrt{\frac{2}{z}}k_F$, where $k_F $ is
the free-electron Fermi wave vector, and z is the nominal valence
\cite{Ashcroft}.
\\
In the presence of scatterings, the non-equilibrium distribution
function will be derived by solving the Boltzmann equation (in
steady state for a homogeneous system),
\begin{equation}
\label{kdot} \dot{\bf{k}}.\frac{\partial f_\lambda}{\partial \bf
k}=(\frac{\partial f_\lambda}{\partial t})_{coll},
\end{equation}
where $\dot{{\bf k}}=\frac{-e \bf E}{\hbar}$ and $(\frac{\partial
f_\lambda}{\partial t})_{coll}$ is called the collision integral,
that in elastic scattering approximation reads \cite{Ramayya}
  \begin{eqnarray}
(\frac{\partial f_\lambda}{\partial t})_{coll}&=& -\sum_{{\bf
k}'q'\lambda'}W_{{\bf k}'\lambda'n\acute{q},{\bf k}\lambda n_q}
 f_\lambda({\bf k},q)(1-f_{\lambda'}({\bf
k}',q'))\delta(\epsilon_{{\bf
k}\lambda}-\epsilon_{{\bf k}'\lambda'}) \nonumber\\
&&+\sum_{{\bf k}'q'\lambda'}W_{{\bf k}'\lambda'n\acute{q},{\bf
k}\lambda n_q} f_{\lambda'}({\bf k}',q')(1-f_\lambda({\bf k},q))
\delta(\epsilon_{{\bf k}\lambda}-\epsilon_{{\bf k}'\lambda'}).
\end{eqnarray}
In this equation, $W_{{\bf k}'\lambda'n\acute{q},{\bf k}\lambda
n_q}$ are the transition probabilities that are given by the Fermi's
golden rule, $W_{{\bf k}'\lambda'n\acute{q},{\bf k}\lambda
n_q}=\frac{2\pi}{\hbar}|V_{{\bf k}'\lambda'n\acute{q},{\bf
k}\lambda n_q}|^2$. \\
\\
Since $\delta_{n'_q,n_q}$ selects only the diagonal elements of
$|V_{{\bf k}'\lambda'n\acute{q},{\bf k}\lambda n_q}|^2$ while
$\delta_{n'_q,n_q-1}\delta_{{\bf k}',{\bf k}+{\bf q}}$ and
$\delta_{n'_q,n_q+1}\delta_{{\bf k}',{\bf k}-{\bf q}}$ select some
of the non-diagonal elements of $|V_{{\bf k}'\lambda'n\acute{q},{\bf
k}\lambda n_q}|^2$, therefore one can easily obtain
\begin{eqnarray}
|V_{{\bf k}'\lambda'n\acute{q},{\bf k}\lambda n_q}|^2=|<{{\bf
k}'}{\lambda'}{{n'}_q}\mid {\hat{H}}_{el-ph}\mid {\bf k}\lambda
n_q>|^2+|<{{\bf k}'}{\lambda'}{{n'}_q}\mid {V}_{im}\mid {\bf
k}\lambda n_q>|^2,
\end{eqnarray}
and accordingly
\begin{eqnarray}
W_{{\bf k}'\lambda'n\acute{q},{\bf k}\lambda n_q}=W^{(1)}_{{\bf
k}'\lambda'n\acute{q},{\bf k}\lambda n_q}+W^{(2)}_{{\bf
k}'\lambda'n\acute{q},{\bf k}\lambda n_q}+W^{(3)}_{{\bf
k}'\lambda'n\acute{q},{\bf k}\lambda n_q},
\end{eqnarray}
Where $W^{(1)}_{{\bf k}'\lambda'n\acute{q},{\bf k}\lambda n_q}$
comes from the impurity potential,
\begin{eqnarray}
W^{(1)}_{{\bf k}'\lambda'n\acute{q},{\bf k}\lambda n_q}= \left(
\begin{array}{cccc}
0 && J^2\\
J^2 && 0\\
\end{array}
\right)n_i\delta_{n'_q,n_q}.
\end{eqnarray}
In which we have used the following approximation,
$(1/L_xL_y)\sum_{j}\sum_{j'}\exp(i(\vec{k'}-\vec{k}).(\vec{r}_j-\vec{r}_{j'}))=n_i$,
where $n_{i}$ is impurity density, $L_x$, $L_y$ are system
dimensions and it should be noted that the summations have been
performed over the random position of impurities.
\\
Meanwhile $W^{(2)}_{{\bf k}'\lambda'n\acute{q},{\bf k}\lambda n_q}$
and $W^{(3)}_{{\bf k}'\lambda'n\acute{q},{\bf k}\lambda n_q}$ are
representing the electron-phonon interaction, where $ W^{(2)}_{{\bf
k}'\lambda'n\acute{q},{\bf k}\lambda n_q} $ presents phonon
absorption contribution,
\begin{eqnarray}
W^{(2)}_{{\bf k}'\lambda'n\acute{q},{\bf k}\lambda
n_q}&=&\delta_{n'_q,n_q-1}\delta_{{\bf k}', {\bf k}+{\bf
q}}\delta_{\lambda',\lambda}\left(
\begin{array}{cccc}
c_q\sqrt{n_q} && 0\\
0 && c_q\sqrt{n_q}\\
\end{array}
\right),
\end{eqnarray}
and $W^{(3)}_{{\bf k}'\lambda'n\acute{q},{\bf k}\lambda n_q}$ should
be considered for the case of emission,
\begin{eqnarray}
W^{(3)}_{{\bf k}'\lambda'n\acute{q},{\bf k}\lambda
n_q}&=&\delta_{n'_q,n_q+1}\delta_{{\bf k}' ,{\bf k}-{\bf
q}}\delta_{\lambda',\lambda} \left(
\begin{array}{cccc}
c^{*}_q\sqrt{n_q+1} &0\\
0&c^{*}_q\sqrt{n_q+1}\\
\end{array}
\right).
\end{eqnarray}
The energy dispersion of conduction electrons becomes anisotropic in
the presence of the spin-orbit interactions. This anisotropy
manifests itself in the scattering process and one can choose an
anisotropic solution to the Boltzmann equation as follows
\cite{Zhian,Loss1},
\begin{eqnarray}
\label{deltaf} \delta f_\lambda(\bf k, \bf q)=e\frac{\partial
f_0(\epsilon_{{\bf k}\lambda})}{\partial\epsilon_{{\bf
k}\lambda}}[a_{{\bf k} q \lambda}({\bf E}.~{\bf v}_{{\bf k}\lambda})
+b_{{\bf k} q \lambda}({\bf E}\times \hat{e}_z).{\bf v}_{{\bf
k}\lambda}].
\end{eqnarray}
\begin{equation}
{\bf v}_{{\bf k}\lambda}=\frac{1}{\hbar}\nabla_{\bf k}\epsilon_{{\bf
k}\lambda},
\end{equation}
where ${\bf v}_{{\bf k}\lambda}$ is velocity of conduction
electrons, $\delta f_\lambda=f_\lambda-f_0$, $\hat{e}_z$ is a unit
vector perpendicular to the two-dimensional plane, $a_{{\bf k}q
\lambda}$ and $b_{{\bf k}q \lambda}$ are two unknown coefficients
that can be determined self-consistently by using eq. (\ref{kdot})
and eq. (\ref{deltaf}) in which $b_{{\bf k} q \lambda}({\bf E}\times
\hat{e}_z).{\bf v}_{{\bf k}\lambda}$ arises due to the anisotropic
nature of the system. Then one can find that the unknown
coefficients $a_{{\bf k}q \lambda}$ and $b_{{\bf k}q \lambda}$
should satisfy following equations \cite{Zhian}
\begin{equation}
\label{a1} \frac{a_{{\bf k}q \lambda}}{\tau^{(1)}_{{\bf k}
q\lambda}}+\frac{b_{{\bf k}q \lambda}}{\tau^{(2)}_{{\bf
k}q\lambda}}=1,
\end{equation}
\begin{equation}
\label{a2} \frac{a_{{\bf k}q \lambda}}{\tau^{(2)}_{{\bf
k}q\lambda}}-\frac{b_{{\bf k}q \lambda}}{\tau^{(1)}_{{\bf
k}q\lambda}}=0,
\end{equation}
From Eqs. (\ref{a1}) and (\ref{a2}), one can easily obtain
\begin{equation}
a_{{\bf k} q \lambda}=\frac{\tau^{(1)}_{{\bf
k}q\lambda}}{1+[\frac{\tau^{(1)}_{{\bf k}q\lambda}}{\tau^{(2)}_{{\bf
k}q\lambda}}]^2}, ~~~~~b_{{\bf k}q\lambda}=\frac{\tau^{(2)}_{{\bf
k}q\lambda}}{1+[\frac{\tau^{(2)}_{{\bf k}\lambda}}{\tau^{(1)}_{{\bf
k}q\lambda}}]^2},
\end{equation}
\\
in which $\tau^{(1)}_{{\bf k}q\lambda}$ and $\tau^{(2)}_{{\bf
k}q\lambda}$ are two relaxation times, defined by
\begin{eqnarray}
\frac{1}{\tau^{(1)}_{{\bf k}q\lambda}}=\sum_{{\bf k'}q',\lambda'}
W_{{\bf k}'\lambda'n\acute{q},{\bf k}\lambda n_q}\{1-\frac{|{\bf
v}_{{\bf k}'\lambda'}|}{|{\bf v}_{{\bf
k}'\lambda'}|}\cos[\theta({\bf v}_{{\bf k}\lambda}\wedge{\bf
v}_{{\bf k}'\lambda'})]\},
\end{eqnarray}
\begin{eqnarray}
\frac{1}{\tau^{(2)}_{{\bf k}q\lambda}}=\sum_{{\bf k'}q',\lambda'}
W_{{\bf k}'\lambda'n\acute{q},{\bf k}\lambda n_q} \frac{|{\bf
v}_{{\bf k}'\lambda'}|}{|{\bf v}_{{\bf
k}'\lambda'}|}\sin[\theta({\bf v}_{{\bf k}\lambda}\wedge{\bf
v}_{{\bf k}'\lambda'})],
\end{eqnarray}
\\
where $\theta({\bf v}_{{\bf k}\lambda}\wedge{\bf v}_{{\bf
k}'\lambda'})$ is the angle between ${\bf v}_{{\bf k}\lambda}$ and
${\bf v}_{{\bf k}'\lambda'}$.
\\
 The spin current operator is defined
as \cite{Rashba}
\begin{equation}
\hat{J}^i_x=\frac{\hbar}{2}\{\sigma_i,\hat{v}_x\},
\end{equation}
where $\hat{v}_x=\hbar^{-1}(\frac{\partial \hat{H}}{\partial k_x})$
is known as velocity operator. Expectation value of spin-current in
a given scattering state i.e. $J^i_x({\bf k},\lambda)=<{\bf
k}\lambda n_q \mid \hat{J}^i_x \mid {\bf k}\lambda n_q>_{scat}$ is
as follows
\begin{eqnarray}
\label{j} J^i_x({\bf k},\lambda)&=&J^{i(0)}_x({\bf k},\lambda)+
\hbar\sum_{{\bf k}'q'\lambda'} Re[V_{{\bf
k}'\lambda'\acute{n}_q,{\bf k}\lambda n_q} <J^i_x>_{{\bf
k}'\lambda'\acute{n}_q,{\bf k}\lambda
n_q} Pr\frac{1}{\epsilon_{{\bf k}\lambda}-\epsilon_{{\bf  k}'\lambda'}}\nonumber\\
&&-V_{{\bf k}'\lambda'\acute{n}_q,{\bf k}\lambda n_q} <J^i_x>_{{\bf
k}'\lambda'\acute{n}_q,k\lambda
n_q}i\pi\delta(\epsilon_{k\lambda}-\epsilon_{k'\lambda'})].
\end{eqnarray}
\\In which we have defined $J^{i(0)}_x= <{\bf k}\lambda n_q \mid
\hat{J}^i_x \mid {\bf k}\lambda n_q>$. Then the transport
spin-current in x direction with spin parallel to the x or y axes,
is given by
\begin{equation}
J^i_x=\sum_{{\bf k}q,\lambda}\hat{J}^i_x({\bf k},\lambda)\delta
f_\lambda({\bf k, \bf q}),~~~~~~~~(i=s_x,s_y)
\end{equation}
\section{Results}
\begin{figure}[htbp]
\centering
  \includegraphics[width=4.0in]{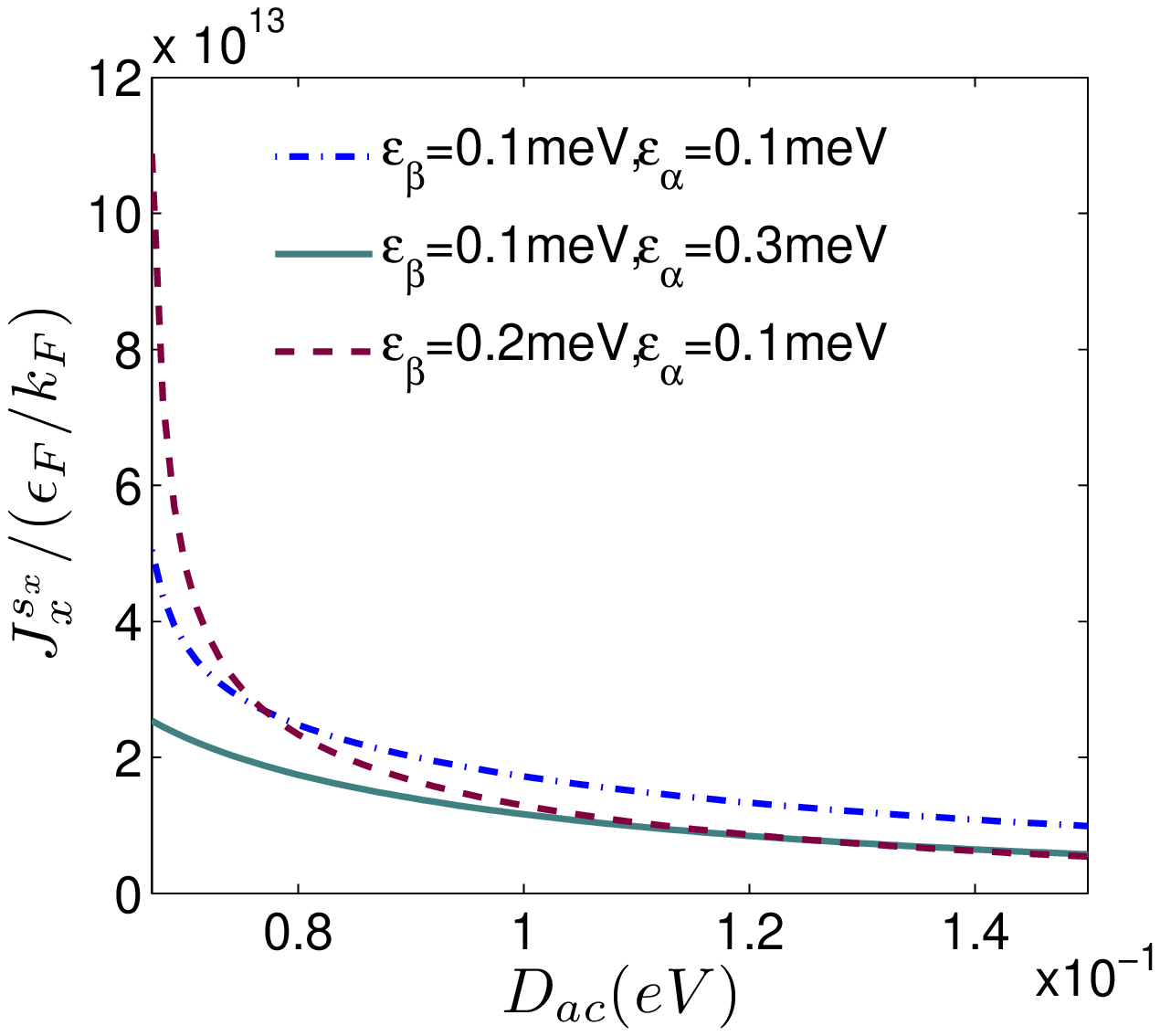}
  \caption{
  %Longitudinal spin current
%as a function of the deformation potential for different SO
%couplings.
}
 \label{fig1}
\end{figure}

\begin{figure}[htbp]
\centering
  \includegraphics[width=4.0in]{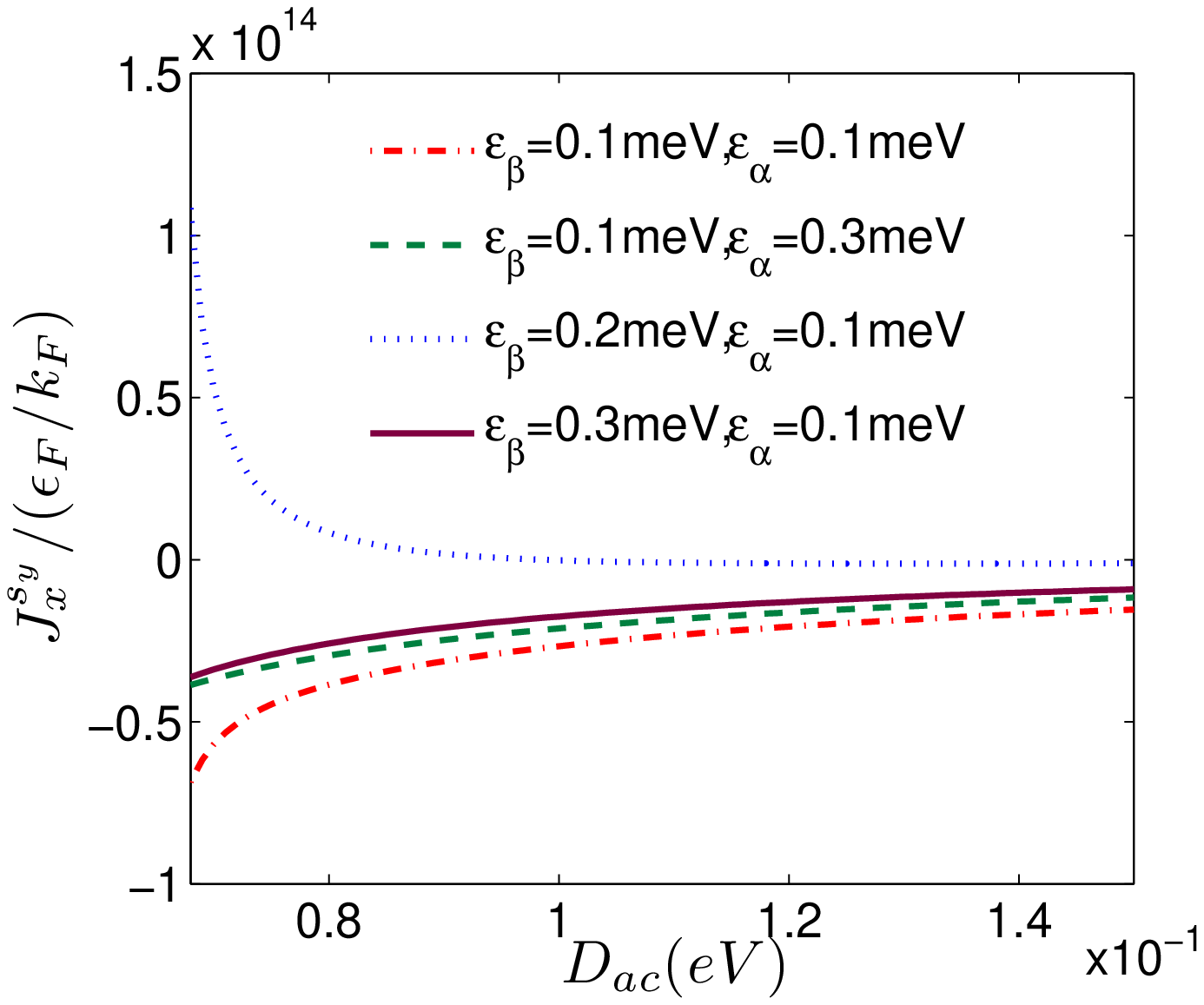}
  \caption{
  %Transverse spin current
%as a function of the deformation potential for different SO
%couplings.
}
\label{fig2}
\end{figure}

\begin{figure}[htbp]
\centering
  \includegraphics[width=4.0in]{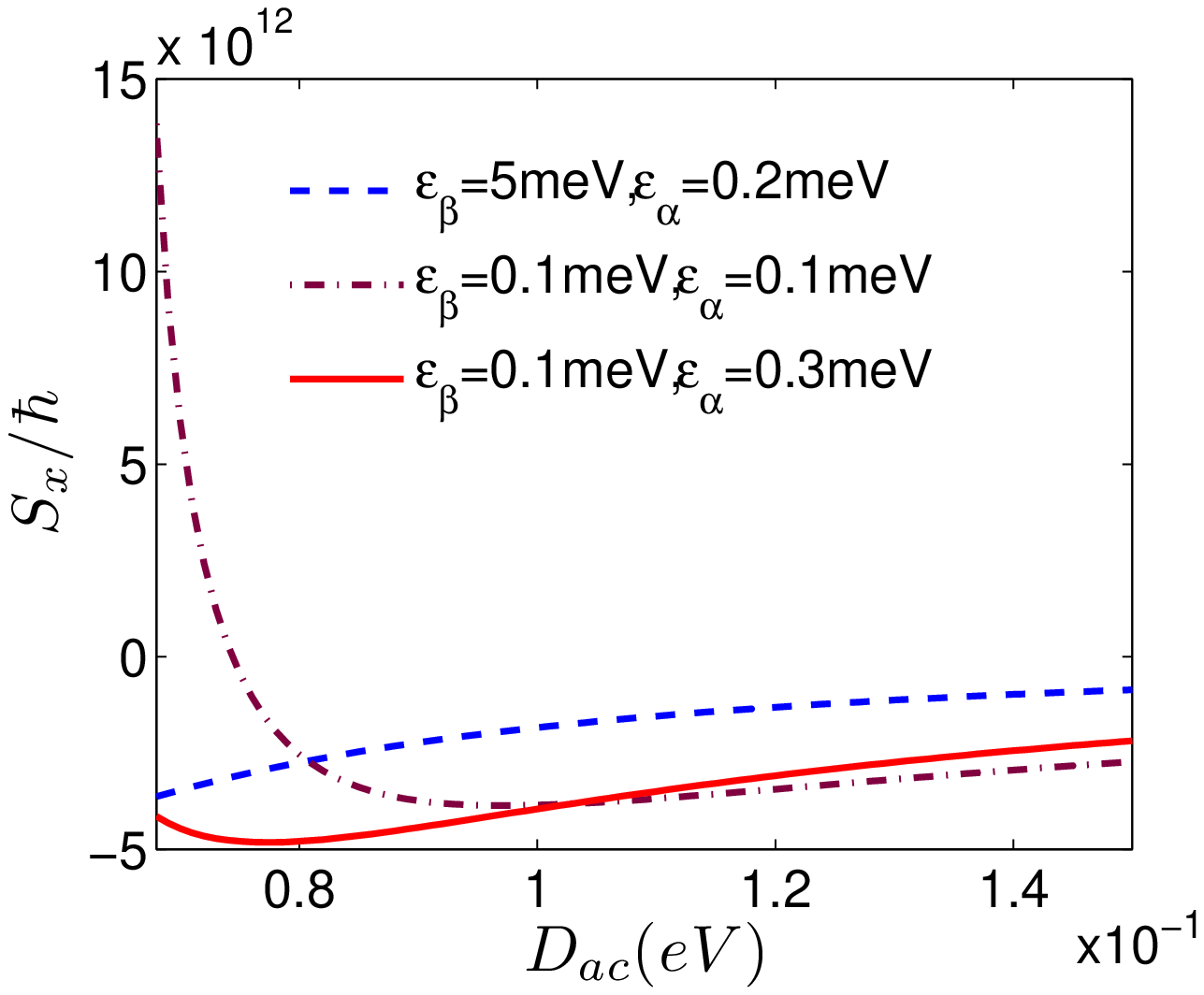}
  \caption{
  %Longitudinal spin accumulation
%as a function of the deformation potential for different SO
%couplings.
}
\label{fig3}
\end{figure}

\begin{figure}[htbp]
\centering
  \includegraphics[width=4.0in]{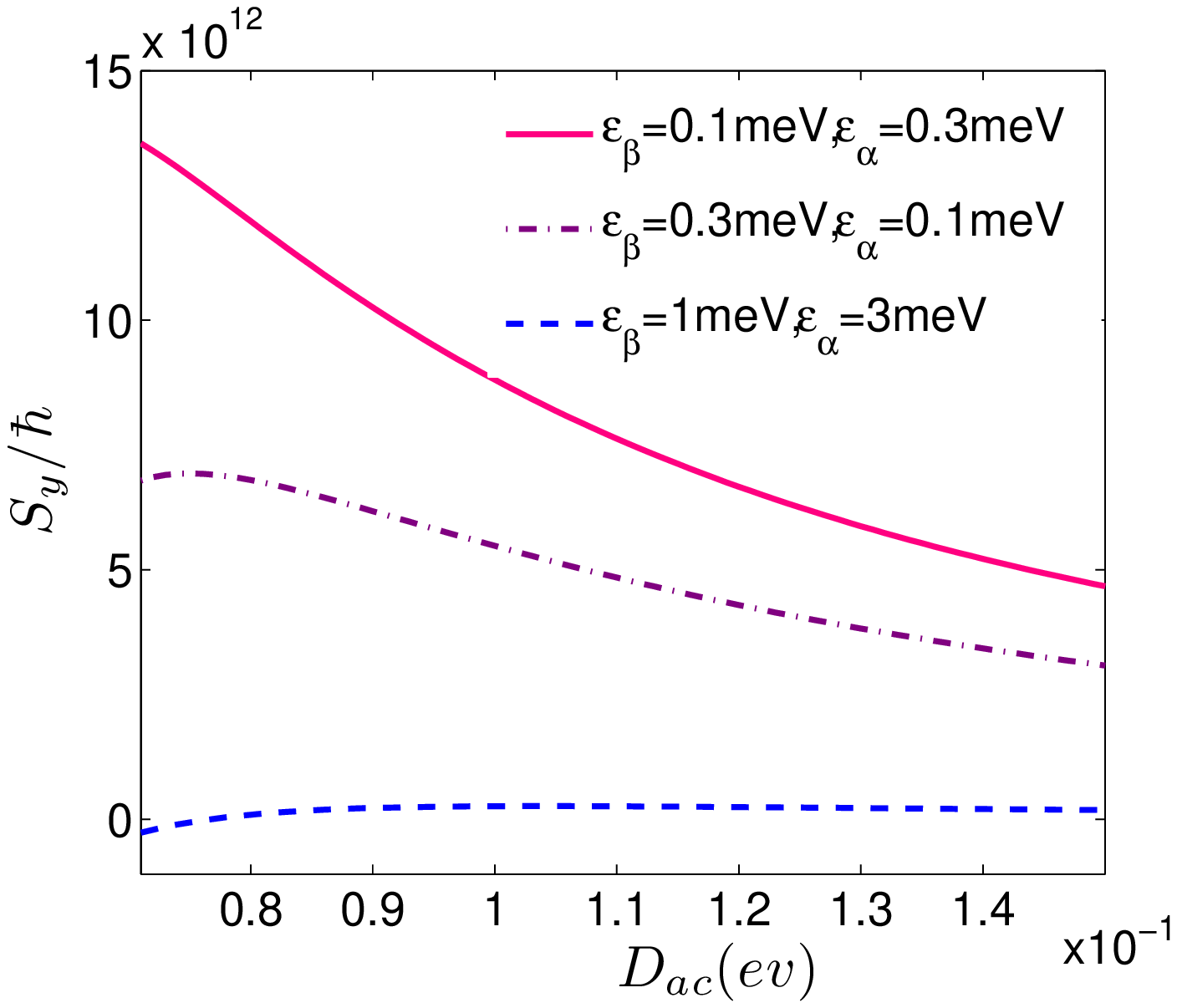}
  \caption{
  %Transverse spin accumulation
%as a function of the deformation potential for different SO
%couplings.
}
\label{fig4}
\end{figure}

In the current letter, the spin accumulation and spin-current of a
two dimensional electron gas have been obtained in the presence of
the Rashba, Dresselhaus and electron-phonon interactions. This was
accomplished by utilizing a semi-classical model developed for
anisotropic systems. This anisotropy is induced by spin-orbit
couplings in the scattering matrix or in the energy dispersion. As
mentioned before, the Rashba and Dresselhaus couplings cannot be
responsible for generation of spin-current in non-equilibrium regime
\cite{Zhian,Inue}. Meanwhile, non-equilibrium spin accumulation is
effectively controlled by these spin-orbit interactions
\cite{Zhian}. Results of the present letter show that the
electron-phonon interaction has a considerable role in the
generation of spin-current which was expected to be obtained by
spin-orbit couplings. Spin accumulation of the system is also
controlled by the strength of electron-phonon coupling.
\\
In the present system, electric field was assumed to be applied
along the $x$ direction and the numerical parameters have been
chosen as follows $\epsilon_f=10eV$ is the Fermi energy, $J=0.1eV$,
$n_i=10^{10}cm^{-2}$ is the density of impurities, $T=1K$ and
$V_s=4950m/s$. In addition Rashba and Dresselhaus couplings have
been denoted by $\epsilon_\alpha=m\alpha^2/\hbar^2$,
$\epsilon_\beta=m\beta^2/\hbar^2$, respectively. The Rashba coupling
can reach high values up to 0.2 eV in epitaxial graphene grown on a
Ni(111) substrate \cite{Dedkov}. However, in the present letter, a
typically lower range has been chosen for SOC as reported for other
materials.
\\
It has been shown that the electron-phonon interaction could not be
considered as an underestimating effect on spin-dependent
mechanisms. It was also demonstrated that, at low electron-phonon
coupling strengths, the lattice vibrations are more effective.
\\
In Fig. \ref{fig1} longitudinal and  Fig. \ref{fig2} transverse
spin-current has been depicted as a function of the deformation
potential. Spin-current has been induced due to the lattice-electron
interactions. These figures clearly show that spin-current of the
system has a accountable value, in which its sign and magnitude
could be controlled by the SOCs. At the same time, as shown in Figs.
\ref{fig3} and \ref{fig4}, longitudinal and transverse spin
accumulations can be effectively changed by the Rashba and
Dresselhaus couplings.
\\
Therefore these results show that when effect of electron-phonon
interaction is taken into account, in the semi-classical regime, it
turns out that both components of the spin-current could take
non-zero values. It seems that details of the scattering potential
has an important role in the generation of spin-current in the
presence of the Rashba and Dresselhaus couplings. As reported in
\cite{Zhian}, Huang and Hu found that short-range delta function
impurity scatterings (which  actually have spherical symmetry)
result in zero spin-polarized current in the system.
\\
An important feature which can be inferred from the results is the
fact that the absolute value of spin-current and spin accumulation
decreases for high electron-phonon couplings, as depicted in Fig.
\ref{fig1}- Fig. \ref{fig4}. Spin-current induced by the lattice
longitudinal vibrations disappears in the limit of high deformation
potential and rapidly increases for low electron-phonon couplings.
Unlike the spin-orbit couplings, electron-phonon interaction could
change the order of magnitude of the spin-current. However, it
should be noted that numerical results reveals that, in the limit of
$D_{ac} \rightarrow 0$ spin current vanishes abruptly, and it was a
numerical discontinuity (not included in the figures) at $D_{ac}=0$.
If identical conditions are chosen from \cite{Zhian} i.e. when
$D_{ac}=0$ and in the case of nonmagnetic impurities, $J=0$,
numerical results show that spin-current identically vanishes which
is in agreement with the results that have been pointed out in
\cite{Zhian} for identical conditions.
\\
As mentioned before, anisotropic effects can be induced by two
different sources: the energy dispersion relation and the scattering
matrix of the anisotropic relaxations. In the current case,
according to the present calculations, the first source of
anisotropy is small and could be neglected. For isotropic
spin-independent relaxations, the scattering matrix should be an
isotropic function, even for anisotropic eigen-state spinors.
However, in the current case of electron-phonon scatterings, it
seems that redistribution of the carriers' population by this
relaxation mechanism could change the ensemble average of Rashba and
Dresselhaus k-dependent effective magnetic field; i.e., in the
presence of this q-dependent relaxation, where $\textbf{q}$ is the
phonon wave number, $\delta f_\lambda(k,q)$ could no longer exhibit
odd function properties. It should be noted that this property
($\delta f_\lambda(-k)=-\delta f_\lambda(k)$) was shown to be
responsible for zero spin current in the system \cite{Zhian}.
%This effective field is given by $\textbf{B}_{eff}^R=1/\mu_B[
%(-\alpha_R k_y+\beta k_x)\hat{x}+ (\alpha_R k_x-\beta k_y)\hat{y}]$.
Meanwhile, this new distribution function, $\delta f_\lambda(k,q)$,
can produce non-vanishing spin current in the system since the
effective field in the present case has been modified. However,
increasing the electron-phonon coupling strength ultimately
decreases both spin current and spin accumulation of the system, as
shown in Fig. \ref{fig1}- Fig. \ref{fig4}. Unlike the intermediate
range of deformation potential at high electron-phonon couplings,
the momentum of electrons is effectively randomized by the
electron-phonon interaction since the relaxation time of the states
decreases. Therefore in this case, population of the carriers
approaches to the limit of isotropic scatterings, in which the spin
current of the system vanishes.
\section{Conclusion}
In this letter, a semi-classical approach have been implemented for
studying magnetoelectric effects of a 2DEG system. The primary focus
of this letter was on showing that the electron-phonon coupling has
an important role in generation of the spin-current in
non-equilibrium regime since it was verified that the Rashba and
Dresselhaus couplings (when the electron-phonon coupling is absent)
cannot be responsible for spin-current in this regime. It was
numerically verified that, even at low electron-phonon couplings,
the lattice vibrations can mediate in the spin-transport process,
modulated by spin-orbit interactions.
\\
\section{Acknowledgment}
This research was supported by a research fund Number 217/D/1288
from Azarbaijan Shahid Madani University.
\bibliographystyle{model1a-num-names}
\bibliography{<your-bib-database>}

\newpage
Fig. 1: Longitudinal spin current as a function of the deformation
potential for different SO
couplings.
\\
\\

Fig. 2: Transverse spin current as a function of the deformation
potential for different SO couplings.
\\
\\

Fig. 3: Longitudinal spin accumulation as a function of the
deformation potential for different SO couplings.
\\
\\

Fig. 4: Transverse spin accumulation as a function of the
deformation potential for different SO couplings.
 \newpage

\end{document}